\documentclass[11pt]{JHEP3}

\newcommand{\N}{{\scriptscriptstyle N}}

\newcommand{\half}{{{\textstyle\frac{1}{2}}}}
\newcommand{\quarter}{{{\textstyle\frac{1}{4}}}}

\newcommand{\be}{\begin{equation} }
\newcommand{\ee}{\end{equation} }
\newcommand{\ba}{\begin{array}}
\newcommand{\ea}{\end{array}}

\newcommand{\su}{\mbox{su}}
\newcommand{\so}{\mbox{so}}
\newcommand{\SO}{\mbox{SO}}

\def\E{{\cal E}}
\def\L{{\cal L}}
\def\P{{\cal P}}
\def\R{{\cal R}}
\def\A{{\cal A}}
\def\B{{\cal B}}

\def\hpartial{\hat{\partial}}
\def\g{g_{\scriptscriptstyle
{Y\!M}}}

\def\tr{{\rm tr}}
\def\Tr{{\rm Tr}}
\def\trN{{\rm tr}_{\!{\scriptscriptstyle N}}}

\def\I_M{{I_{\scriptscriptstyle M\times M}}}

\def\N{{\cal  N}}

\title{5D action for longitudinal five  branes on a pp-wave}

\author{Seungjoon Hyun and Jeong-Hyuck Park\footnote{\textsl{On leave from}
Korea Institute for   Advanced Study, Seoul.}\\
Institute of Physics and Applied Physics, Yonsei University, Seoul 120-749, Corea\\
Electronic correspondence : \email{hyun@phya.yonsei.ac.kr},
\email{jhp@kias.re.kr}}

\abstract{String modes in a pp-wave background are generically
massive, and the worldvolume description  of the branes is to be
given by  `massive' gauge theories. In this paper, we present a
five dimensional  super Yang-Mills action with the
K\"{a}hler-Chern-Simons term plus  the Myers term  as a
low energy worldvolume description of the longitudinal five  branes in a
maximally  supersymmetric pp-wave background. We derive the action
from the M-theory  matrix model on the pp-wave. We utilize the
previously found $4/32$ BPS solution of rotating  five  branes with
stacks of membranes, but,  to obtain the static configuration, we
reformulate the matrix model in a rotating coordinate system
which provides the inertial frame for the branes. Expanding the
matrix model around the solution, we first obtain a
non-commutative field theory action naturally equipped with the
full sixteen dynamical supersymmetries. In the commutative limit,
we show only four supersymmetries survive, resulting in a novel five dimensional
``${\cal N}=1/2$" theory. }

\keywords{Longitudinal five  brane, pp-wave, supersymmetry}

\preprint{KIAS-P02058\\
hep-th/0209219}

\begin{document}

\section{Introduction}

Recently \cite{Berenstein:2002jq},    Berenstein, Maldacena and
Nastase (BMN) proposed a novel
matrix model which describes M-theory in  the maximally  supersymmetric
pp-wave background of the eleven dimensional supergravity
\cite{Kowalski-Glikman,Figueroa-O'Farrill1,FO2},
\begin{equation}
\ba{l}ds^{2}=-2dx^{+}dx^{-}-\Big[(\textstyle{\frac{\mu}{6}})^{2}(x_{1}^{2}+\cdots+x_{6}^{2})+
(\textstyle{\frac{\mu}{3}})^{2}(x_{7}^{2}+x_{8}^{2}+x_{9}^{2})\Big]dx^{+}dx^{+}
+\displaystyle{\sum_{A=1}^{9}}\,dx^{A}dx^{A}\,,\\
{}\\
F_{+789}=\mu\,, \ea\label{pp-wave}
\end{equation}
where $\mu$ becomes the  characteristic   mass parameter of the matrix model.
The resulting matrix model  corresponds to a mass deformation of the BFSS matrix model
\cite{Banks:1996vh,Susskind:1997cw,Sen,Seiberg}, still maintaining the full supersymmetries,
sixteen dynamical and sixteen kinematical.
The BMN matrix model was also shown to agree with  the matrix regularization \cite{deWit,Hoppe}
of the supermembrane on the pp-wave geometry \cite{Dasgupta:2002hx}. \\

Due to   the mass parameter, the BMN  matrix model captures many
interesting novel properties. The supersymmetry transformations
have the explicit time dependence so that the supercharges do not
commute with the Hamiltonian. As a result, the bosons and
fermions have different masses. The bosonic mass terms lift up
the flat directions completely, and the perturbative expansion is
possible by powers of the dimensionless parameter, $(\mu
l_{p}^{2}/R)^{-1}$, where $l_{p}$ is the eleven dimensional Planck
length and $R$ is the radius of the null compactification
\cite{Dasgupta:2002hx,Kim:2002if}. Classical vacua are given by
fuzzy spheres sitting at the origin stretching
over the $7,8,9$ directions.\\

In \cite{JhpMsusy} (see also \cite{Keshav}),  the  supersymmetry
algebra of the BMN matrix model was identified as the special
unitary Lie superalgebra of which  the complexification
corresponds  to $\mbox{A}(1|3)$, and   the classification of the
quantum BPS multiplets was carried out as its atypical
representations. Soon after, in \cite{Park:2002cb}, the classical
counterparts of the quantum BPS states were studied. Namely,  all
the BPS equations which correspond to the quantum BPS states
preserving some fraction of the dynamical supersymmetry  were
obtained. The results show that there are essentially one unique
set of $2/16$ BPS equations, three inequivalent sets of $4/16$
BPS equations, and three inequivalent sets of $8/16$ BPS
equations only, in addition to the $16/16$ static fuzzy sphere.
The solutions include the known ones,
 rotating longitudinal five  branes with stacks of D2 branes
in them \cite{Hyun:2002cm}, rotating ellipsoidal branes, rotating
or static hyperboloids \cite{Bak:2002rq}, rotating fuzzy torus
\cite{Mikhailov:2002wx}, and also new  ones such as the rotating
fuzzy spheres or D0 branes in various directions with different
supersymmetries, a static
fuzzy sphere on  a hyperboloid, a mixture of  rotating two hyperboloids
and a fuzzy sphere \cite{Park:2002cb}.\\

Especially, among them the   solution describing  rotating
longitudinal five  branes with stacks of D2 branes is of particular
interest in the present paper. From the classification of the BPS
equations it appears that  the solution  is the unique `flat'
longitudinal five  brane solution which preserves only the dynamical
supersymmetries. The configuration satisfies `the $\mbox{su}(2)$
singlet $4/16$ BPS equations' so that it preserves four dynamical
supersymmetries only. This contrasts to the BFSS matrix model or
$\mu=0$ case where the longitudinal five  brane with stacks of D2
branes preserves half of thirty
two supersymmetries. More detailed comparison is given later.\\

It is also worth to note that  there are supersymmetric
configurations which  preserve only certain  nontrivial
combinations of the dynamical and kinematical supersymmetries.
They include  a transverse membrane and a longitudinal five   brane
\cite{Hyun:2002cm}. Since the kinematical supercharges and the
dynamical supercharges in the BMN matrix model have different
quantum numbers for the Hamiltonian,  such configurations do
not correspond to the energy eigenstates but rather superpositions. \\

One characteristic feature of the string theory in a pp-wave
background is that the string modes are generically massive \cite{Metsaev:2001bj,Metsaev:2002re,Alishahiha:2002nf,Hyun:2002wu},

\begin{equation}
E_{n}=\sqrt{\mu^{2}+n^{2}/(\alpha^{\prime}p_{+})^{2}}\,.
\end{equation}
Therefore, in the $\alpha^{\prime}\rightarrow 0$ limit,  the
worldvolume descriptions of the branes are to be given by
`massive' gauge theories.\footnote{An attempt to build such field
theories was taken in \cite{Bonelli:2002mb}.} \\

In this paper,  we present a five dimensional super Yang-Mills
action with  the K\"{a}hler-Chern-Simons term plus the Myers
term  as a low energy worldvolume description of the longitudinal five  branes
in a maximally supersymmetric  pp-wave background. We derive the
action in the M-theory matrix model  setup. We utilize the known
 BPS solution of rotating  five  branes with stacks of transverse membranes
or D2 branes, but, to obtain the static configuration, we reformulate
the BMN matrix model in a rotating coordinate system which
provides the inertial frame for the branes.
 The modified matrix model naturally admits  flat and static longitudinal
five  branes with
stacks of D2 branes in them which preserve four dynamical
supersymmetries. We first expand the modified matrix model around
the solution, and  obtain a non-commutative field theory
naturally equipped with the full sixteen supersymmetries. Taking
the commutative limit and letting the D2 branes disappear,  we
finally get the worldvolume action for the longitudinal
five  . We show only four
supersymmetries survive, resulting in a novel five dimensional
``${\cal N}=1/2$" theory.\\

The organization of the present paper is as follows. In section
\ref{setup},  we first reformulate  the BMN matrix model by
introducing a new coordinate system. In this setup, we identify
the BPS equations for the supersymmetric  configurations which
preserve four dynamical supersymmetries, and as a special solution
we find flat and static longitudinal five  branes with stacks of D2
branes in them.  Expanding the matrix model around the solution
we derive a non-commutative five dimensional $\mbox{U}(N)$ super
Yang-Mills action with  the K\"{a}hler-Chern-Simons term plus the
Myers term equipped with the full sixteen dynamical
supersymmetries. In section \ref{WV} we take the commutative
limit to obtain the worldvolume action for  the longitudinal five
branes on the pp-wave. The D2 branes are now gone and the
resulting commutative action has only four supersymmetries. We
study the supersymmetry algebra and identify the central and
$R$-symmetry charges. We consider the BPS configurations which
preserve all the four supersymmetries and write the corresponding
BPS equations. We also discuss the energy spectra of the bosons
and fermions, and show that the five dimensional  ${\cal N}=1/2$ model contains
three supermultiplets. Finally, in section \ref{conclusion}, we
conclude with the summary. The appendix contains  some useful
formulae.

\newpage

\section{M-theory matrix model on a fully
supersymmetric pp-wave\label{setup}}
\subsection{BMN matrix model in the rotating  coordinate system}
The original BMN matrix model or the M-theory matrix model on a
fully  supersymmetric pp-wave background admits the rotating flat
longitudinal five  branes as a BPS solution preserving four
supersymmetries \cite{Hyun:2002cm,Park:2002cb}. For the purpose of
the present paper, we choose the comoving or  inertial coordinate
system such that  the longitudinal five  brane solution becomes
static. Explicitly we replace the first four coordinates,
$x_{1},x_{2},x_{3},x_{4}$, by the $\SO(2)\times\SO(2)$ rotating
ones,
\begin{equation}
\ba{cc} x_{1}\rightarrow \cos(\mu x^{+}/6)x_{1}+\sin(\mu x^{+}/6)x_{2}\,,~~~&~~~x_{2}\rightarrow
\cos(\mu x^{+}/6)x_{2}-\sin(\mu x^{+}/6)x_{1}\,,\\
{}&{}\\
x_{3}\rightarrow \cos(\mu x^{+}/6)x_{3}+\sin(\mu x^{+}/6)x_{4}\,,~~~&~~~
x_{4}\rightarrow \cos(\mu x^{+}/6)x_{4}-\sin(\mu
x^{+}/6)x_{3}\,, \ea
\end{equation}
so that the metric of the eleven dimensional  pp-wave
background~(\ref{pp-wave}) is, in the new coordinate system, of
the form
\begin{equation}
\ba{ll}ds^{2}=&-2dx^{+}dx^{-}-\textstyle{\frac{\mu}{3}}(x_{1}dx_{2}-x_{2}dx_{1}+x_{3}dx_{4}-x_{4}dx_{3})dx^{+}\\
{}&{}\\
{}& -\Big[(\textstyle{\frac{\mu}{6}})^{2}(x_{5}^{2}+x_{6}^{2})
+(\textstyle{\frac{\mu}{3}})^{2}(x_{7}^{2}+x_{8}^{2}+x_{9}^{2})
\Big]dx^{+}dx^{+}+\displaystyle{\sum_{A=1}^{9}}\,dx^{A}dx^{A}\,.
\ea\label{pp-wave2}
\end{equation}

The corresponding M-theory matrix model on this background is
then obtained from the original  BMN matrix model by taking the
above time dependent $\SO(2)\times\SO(2)$ rotation. With $t\equiv
x^{+}$,  the transformation of the bosons is essentially the same
as above ({\it cf.} \cite{Sugiyama:2002tf,Hyun:2002wu}),
\begin{equation}
\ba{ll} X_{1}\rightarrow \cos(\mu t/6)X_{1}+\sin(\mu
t/6)X_{2}\,,~~&~~\mbox{etc.}\ea\label{XROTATION}
\end{equation}
while that of the fermions reads, from the standard Lorentz
transformation rule,
\begin{equation}
\Psi~\rightarrow~\displaystyle{e^{\frac{\mu}{12}(\Gamma^{12}+\Gamma^{34})t}\Psi\,.}\label{PSIROTATION}
\end{equation}
\newline

The  modified, but nevertheless equivalent,  M-theory matrix
model on a fully supersymmetric  pp-wave background spells with a
mass parameter, $\mu$,
\begin{equation}
{\cal
S}=\displaystyle{\frac{l_{p}^{6}}{R^{3}}}\displaystyle{\int\,dt}~
\L_{0}+\mu\L_{1}+\mu^{2}\L_{2}\,, \label{Maction}
\end{equation}

\begin{equation}
\ba{l} \L_{0}=\Tr\!\left(\half D_{t}X^{A}D_{t}X_{A}+\quarter
[X^{A},X^{B}]^{2}+i\half
\Psi^{\dagger}D_{t}\Psi-\half\Psi^{\dagger}\Gamma^{A}[X_{A},\Psi]\right)\,,\\
{}\\
\L_{1}=\Tr\!\left[-\textstyle{\frac{1}{6}}J^{ij}X_{i}D_{t}X_{j}
-i\textstyle{\frac{1}{3}}\epsilon^{rst}X_{r}X_{s}X_{t}
+i\textstyle{\frac{1}{24}}\Psi^{\dagger}(\Gamma^{12}+\Gamma^{34}+3\Gamma^{789})\Psi \right]\,,\\
{}\\
\L_{2}=-\,\half\,\Tr\!\left[(\textstyle{\frac{1}{6}})^{2}(X_{5}^{\,2}
+X_{6}^{\,2})+(\textstyle{\frac{1}{3}})^{2}
(X_{7}^{\,2}+X_{8}^{\,2}+X_{9}^{\,2})\right]\,, \ea
\end{equation}
where   $i,j=1,2,3,4$, $~r,s,t=7,8,9$, $~A,B=1,2,\cdots,9$  and
$J^{ij}$ is a skew-symmetric constant two form of which the
non-vanishing components are $J^{12}=J^{34}=1$ only, up to the
anti-symmetric property. In the present paper, we adopt  generic
Euclidean nine dimensional gamma matrices,
$\Gamma^{A}=(\Gamma^{A})^{\dagger}$, $\Gamma^{12\cdots 9}=1$.
Namely we do not adopt the usual real and symmetric Majorana
representation. Accordingly there exits a nontrivial $16\times
16$ charge conjugation matrix, ${C}$,
\begin{equation}
\begin{array}{ll}
(\Gamma^{A}){}^{T}=(\Gamma^{A}){}^{\ast}={C}^{-1}\Gamma^{A}{C}\,,~~~&~~~C=C^{T}=(C^{\dagger})^{-1}\,.
\end{array}
\end{equation}
The spinors, $\Psi$,  satisfy the Majorana condition leaving eight
independent  complex components,
\begin{equation}
\Psi={C}\Psi^{\ast}\,. \label{Majoranacondition}
\end{equation}
The covariant derivatives are in our convention, $D_{t}{\cal O}=\frac{d~}{dt}{\cal O}-i[A_{0},{\cal O}]$ so
that $X$ and $A_{0}$ are of the mass dimension one, while $\Psi$ has the mass dimension ${3}/{2}$.
\newline

Compared to the original BMN matrix model, the quadratic mass
terms for the bosonic first four coordinates are absent. Instead,
there appear terms linear in $\mu$ as well as the velocities.
Consequently the linearly realized isometry group is broken as
\begin{equation}
\SO(6)\times\SO(3)\rightarrow\mbox{SU}(2)\times\SO(2)\times\SO(3)\,,
\end{equation}
which is the price we pay in order  to get the static flat
longitudinal five  brane configurations  we discuss shortly.
\newline

The supersymmetry transformations are
\begin{equation}
\ba{l} \delta A_{0}=i\Psi^{\dagger}\E(t)\,,~~~~~~~~~~~\delta X^{A}=i\Psi^{\dagger}\Gamma^{A}\E(t)\,,\\
{}\\
\delta\Psi=\Big[D_{t}X^{A}\Gamma_{A}-i\half
[X^{A},X^{B}]\Gamma_{AB}+
\textstyle{\frac{\mu}{6}}(X^{5}\Gamma_{5}+X^{6}\Gamma_{6}-2X^{7}\Gamma_{7}-2X^{8}
\Gamma_{8}-2X^{9}\Gamma_{9})\Gamma^{789}\\
{}\\
~~~~~~~~~~~+\textstyle{\frac{\mu}{6}}(X^{1}\Gamma_{1}+X^{2}\Gamma_{2})(\Gamma^{789}-\Gamma^{12})+
\textstyle{\frac{\mu}{6}}(X^{3}\Gamma_{3}+X^{4}\Gamma_{4})(\Gamma^{789}-\Gamma^{34})\Big]\E(t)\,,
\label{susytr}\ea
\end{equation}
where
\begin{equation}
\ba{cc}
\displaystyle{\E(t)=e^{\frac{\mu}{12}(-\Gamma^{12}-\Gamma^{34}+\Gamma^{789})t}\E\,,}~~~&~~~
\E={C}\E^{\ast}\,,\ea \label{Et}
\end{equation}
and $\E$ is a sixteen component constant spinor.\\

In addition there is  the kinematical  supersymmetry,
\begin{equation}
\ba{ccc} \delta A_{0}=\delta X^{A}=0\,, ~~~~&~~~~
\displaystyle{\delta\Psi=e^{-\frac{\mu}{12}(\Gamma^{12}+\Gamma^{34}+3\Gamma^{789})t}\E^{\prime}\,,}
~~~&~~~\E^{\prime}={C}\E^{\prime}{}^{\ast}\,.\ea\label{Eprimet}
\end{equation}
\newpage

\subsection{Static longitudinal five  branes preserving four supersymmetries}
In general, the Killing spinors in the supersymmetry
transformations form a kernel space. Analyzing  `the projection
matrix' to the kernel,   one can obtain in a systematic way all
the possible sets of the BPS equations of various  unbroken
supersymmetry fractions \cite{jhpBPS,Park:2002cb}.  In order to
obtain the static longitudinal five  brane configuration,  it is
convenient  to consider the following $4/16$ projection matrix
for the Killing spinors \cite{Park:2002cb},
\begin{equation}
\Omega=\quarter(1-\Gamma^{1234}-\Gamma^{3456}-\Gamma^{5612})\,,\label{Omega}
\end{equation}
which satisfies
\begin{equation}
\ba{cccc} \Omega^{\dagger}=\Omega\,,~~~&~~~ C\Omega^{\ast}C^{-1}=\Omega\,,~~~&~~~
\Omega^{2}=\Omega\,,~~~&~~~\tr\Omega=4\,.\ea
\end{equation}
Now replacing the Killing spinor, $\E$, in (\ref{susytr}) by the
projection matrix,  rewriting the expression in terms of the
totally anti-symmetric products of gamma matrices and requiring
each coefficient to vanish one can obtain the following  BPS
equations preserving four supersymmetries,
\begin{equation}
\begin{array}{ll}
D_{t}Z_{1}=D_{t}Z_{2}=D_{t}X_{r}=0\,,~~~&~~~D_{t}Z_{3}+i\textstyle{\frac{\mu}{6}}Z_{3}=0\,,\\{}&{}\\
{}[X_{r},X_{s}]-i\textstyle{\frac{\mu}{3}}\epsilon_{rst}X^{t}=0\,,~~~&~~~[X_{r},X_{A}]=0,~A=1,2,\cdots,6\,,\\
{}&{}\\
{}[Z_{1},Z_{2}]=0\,,~~~&~~~[Z_{1},\bar{Z}_{1}]+[Z_{2},\bar{Z}_{2}]+[Z_{3},\bar{Z}_{3}]=0\,,\\
{}&{}\\
{}[Z_{2},Z_{3}]=0\,,~~~&~~~[Z_{3},Z_{1}]=0\,, \ea\label{MBPS}
\end{equation}
where we complexify the coordinates as
$Z_{1}=X_{1}+iX_{2},\,Z_{2}=X_{3}+iX_{4},\,,Z_{3}=X_{5}+iX_{6}$,
and set $\bar{Z}_{1}=(Z_{1})^{\dagger}$ etc. Note that  the BPS
equations themselves imply the Gauss constraint. Rotating back to
the original  coordinates, $Z_{1},Z_{2}\rightarrow e^{i\mu
t/6}Z_{1},e^{i\mu t/6}Z_{2}$, this set of BPS equations is
identical to the $\su(2)$ singlet  BPS equations preserving four
supersymmetries found in  \cite{Park:2002cb}.\\

Generic finite matrix solutions describe  the fuzzy sphere or the
giant graviton expanding in the $7,8,9$ directions and rotating on
the $(5,6)$ plane with the frequency,  $\mu/6$, since the last
four equations imply that $Z_{1},Z_{2},Z_{3}$ are simultaneously
diagonalizable.  On the other hand, for the infinite matrix
solutions, by setting $X_{r}=Z_{3}=A_{0}=0$, one can obtain the
static flat longitudinal  five  branes \cite{Hyun:2002cm},
\begin{equation}
\ba{ccc}
{}[X^{1},X^{2}]+[X^{3},X^{4}]=0\,,~&~[X^{1},X^{3}]+[X^{4},X^{2}]=0\,,~&~
[X^{1},X^{4}]+[X^{2},X^{3}]=0\,.\ea\label{BPSLM5}
\end{equation}
In the present paper,  we consider the longitudinal five  branes with
stacks of D2 branes in them \cite{Banks:1996nn} as solutions,
\begin{equation}
X^{i}=i\hat{\partial}_{i}\,,~~~i=1,2,3,4\,.\label{LM5sol}
\end{equation}
Here $\hat{\partial}_{i}$'s are related to the coordinates of a
four dimensional non-commutative space,
\begin{equation}
x^{i}=i\theta^{ij}\hpartial_{j}\,,
\end{equation}
such that
\begin{equation}
\ba{ccc}
[x^{i},x^{j}]=i\theta^{ij}\,,~~~&~~~[\hpartial_{i},\hpartial_{j}]=i\theta^{-1}_{\,ij}\,,~~~&~~~
[\hpartial_{i},x^{j}]=\delta_{i}{}^{j}\,.\ea\label{non-com}
\end{equation}
In order to satisfy the BPS condition (\ref{BPSLM5}),  the
noncommutative parameter must satisfy the anti-self-duality,
\begin{equation}
\ba{ccc}\theta^{ij}+\half\epsilon^{ijkl}\theta^{kl}=0
~~&~~\Longleftrightarrow~~&~~\theta^{-1}_{\,ij}+\half\epsilon_{ijkl}\theta^{-1}_{\,kl}=0\,.\ea\label{asd}
\end{equation}
The relation (\ref{non-com}) defines a pair of non-commutative
planes, and hence  two sets of the harmonic oscillators. The most
general irreducible representation is then specified by the
superselection rule which is the number of the ground states that
we denote by $N$.  Thus, the Hilbert space, ${\cal H}$, on which
the infinite matrices act decomposes as a direct product of two
harmonic oscillator Hilbert spaces, $H_{h.o.}$ and an $N$
dimensional vector space, $V_{N}$,
\begin{equation}
{\cal H}=H_{h.o.}\oplus H_{h.o.}\oplus V_{N}\,.
\end{equation}
Explicitly as  in \cite{JHPcomments}, using the bra and ket
notation one can  regroup the states in the Hilbert space  as
\begin{equation}
\ba{cccc}
|n_{1},n_{2},s\rangle\,,~~&~~n_{1},n_{2}=0,1,\cdots,\infty\,,~&~s=1,2,\cdots,N\,,
\ea
\end{equation}
so that the two creation operators are
\begin{equation}
\ba{ll}
\displaystyle{\sum_{n_{1},n_{2},s}\sqrt{n_{1}+1}|n_{1}+1,n_{2},s\rangle\langle
n_{1},n_{2},s|\,,}~~&~~
\displaystyle{\sum_{n_{1},n_{2},s}\sqrt{n_{2}+1}|n_{1},n_{2}+1,s\rangle\langle
n_{1},n_{2},s|\,.} \ea
\end{equation}
In terms of branes, this represents $N$ parallel longitudinal five
branes on top of each other with stacks of D2 branes in them,
which preserve four supersymmetries.\\

 It is worth to note that in
the ordinary BFSS matrix model  or the $\mu=0$ case, the same
longitudinal five  brane configuration,
(\ref{LM5sol},\ref{non-com},\ref{asd}),  preserves  sixteen
supersymmetries out of thirty two. They are  eight of the
dynamical supersymmetries with the projection matrix,
$\quarter(1-\Gamma^{1234})$, and  eight linear  combinations of
the kinematical and dynamical supersymmetries, since the remaining
dynamical supersymmetry transformations of the fermions are
canceled by the kinematical supersymmetry transformations.
Furthermore, it is  possible to relax the anti-self-duality
condition (\ref{asd}). In that case, the longitudinal five  brane
configuration preserves sixteen linear combinations of the
kinematical and dynamical supersymmetries. However, in the case of
$\mu\neq 0$, the mixing between the kinematical  and dynamical
supersymmetries is not allowed because of the different time
dependence  in (\ref{Et}) and (\ref{Eprimet}).  In summary, the
flat longitudinal five  branes are  4/32 supersymmetric in the
pp-wave background, while 16/32 supersymmetric in the flat
 background.\newpage

\subsection{Non-commutative 5D super Yang-Mills-K\"{a}hler-Chern-Simons-Myers action}
In this subsection, we  expand our  M-theory matrix model around
the supersymmetric  $N$ parallel longitudinal five  brane solution
above, and derive a five dimensional super Yang-Mills action
coupled to the K\"{a}hler-Chern-Simons term plus  the Myers
term.\\

Introducing the gauge fields as the longitudinal fluctuations
around the five  brane solution, we  write the bosonic variables as
\begin{equation}
\ba{ll}X_{i}=i\hpartial_{i}+A_{i}\,,~~~&~~i=1,2,3,4\,,\\
{}&{}\\
X_{a}=\Phi_{a}\,,~~~&~~a=5,6,7,8,9\,.\ea
\end{equation}
Consequently
\begin{equation}
\ba{ll} D_{t}X_{i}=F_{0i}\,,~~~~&~~~~
{}[X_{i},X_{j}]=i(F_{ij}-\theta^{-1}_{\,ij})\,,\\
{}&{}\\
{}[X_{i},\Phi]=iD_{i}\Phi\,,~~~~&~~~~
{}[X_{i},\Psi]=iD_{i}\Psi\,, \ea
\end{equation}
where
$F_{\mu\nu}=\partial_{\mu}A_{\nu}-\partial_{\nu}A_{\mu}-i[A_{\mu},A_{\nu}]$,
$~\mu,\nu=0,1,2,3,4$ and  the derivative along the non-commutative
coordinate of a function is from (\ref{non-com}),
$\partial_{i}\Phi=[\hpartial_{i},\Phi]$.  The fields have the
standard gauge transformation properties,
\begin{equation}
\ba{ll}A_{\mu}~\rightarrow~UA_{\mu}U^{\dagger}+iU\partial_{\mu}U^{\dagger}\,,~~~~&~~~~\Phi~\rightarrow~U\Phi
U^{\dagger}\,.\ea
\end{equation}

To write the matrix model (\ref{Maction}) in terms of the gauge
fields we first note
\begin{equation}
J^{ij}\Tr(X_{i}D_{t}X_{j})= -\half\epsilon^{\lambda\mu\nu
ij}\Tr(A_{\lambda}\partial_{\mu}A_{\nu}-\textstyle{i\frac{2}{3}}A_{\lambda}A_{\mu}A_{\nu})J_{ij}
+J^{ij}\theta^{-1}_{\,ij}\Tr\,A_{0}+
\displaystyle{\frac{d\,}{dt}\Tr(iJ^{ij}\hpartial_{i}A_{j})\,,}
\end{equation}
where $\epsilon^{\lambda\mu\nu ij}$ is the totally anti-symmetric
five  form tensor  with $\epsilon^{01234}=1$. Now the crucial
observation to make  is that  the second term linear in $A_{0}$ on
the right hand side vanishes due to the anti-self-duality of the
non-commutative parameter, $J^{ij}\theta^{-1}_{\,ij}=0$.
Therefore the right hand side is identified as the
K\"{a}hler-Chern-Simons term \cite{NairKCS} up to the total
derivative with  $J_{ij}$ being the K\"{a}hler form in the
non-commutative flat four dimensional space. Since the left hand
side is manifestly gauge invariant, there will be no quantization
rule for the coefficient of the K\"{a}hler-Chern-Simons term,
contrary to the case in the Chern-Simons
theory on a non-commutative plane \cite{NC-CS}.\\

Now using the fact that the trace over the Hilbert space, ${\cal
H}$, can decompose into  the  integration over the
non-commutative four dimensional space and the trace over the
``$\mbox{U}(N)$'' indices,\footnote{Here we set $\theta^{2}=\mbox{Pfaffian}(\theta^{ij})$.}
\begin{equation}
\displaystyle{\Tr{\cal O}(x)=\frac{1}{(2\pi\theta)^{2}}\int
dx^{4}\,\tr_{\!{\scriptscriptstyle N}}{\cal O}(x)\,,}
\end{equation}
our M-theory matrix model (\ref{Maction}) in the five  brane
background becomes, discarding the total derivative terms and the mass of the five  brane background, a
non-commutative  five dimensional super Yang-Mills action  coupled
to the K\"{a}hler-Chern-Simons term  \cite{NairKCS} plus  the
Myers term \cite{Myers:1999ps},
\begin{equation}
\ba{ll} {\cal S}=\displaystyle{\frac{1}{\g^{2}} \int\,dx^{5}}~
\L_{0}+\mu\L_{1}+\mu^{2}\L_{2}\,,~~~&~~~~\displaystyle{\g^{2}=\frac{(2\pi\theta)^{2}
R^{3}}{l_{p}^{6}}=\frac{(2\pi\theta)^2 g_s}{l_{s}^{3}}}\,,
\ea\label{LM5action}
\end{equation}

\begin{equation}
\ba{l}\L_{0}=\tr_{\!{\scriptscriptstyle N}}\Big[-\quarter
F_{\mu\nu}F^{\mu\nu}-\half D_{\mu}\Phi_{a}D^{\mu}\Phi_{a}+\quarter
[\Phi_{a},\Phi_{b}]^{2}-i\half\Psi^{\dagger}\Gamma^{\mu}D_{\mu}\Psi-\half\Psi^{\dagger}\Gamma^{a}[\Phi_{a},\Psi]\Big]\,,\\
{}\\
\L_{1}=\tr_{\!{\scriptscriptstyle
N}}\Big[\textstyle{\frac{1}{12}}\epsilon^{\lambda\mu\nu
ij}\Tr(A_{\lambda}\partial_{\mu}A_{\nu}-\textstyle{i\frac{2}{3}}A_{\lambda}A_{\mu}A_{\nu})J_{ij}
-i\textstyle{\frac{1}{3}}\epsilon^{rst}\Phi_{r}\Phi_{s}\Phi_{t}
+i\textstyle{\frac{1}{48}}\Psi^{\dagger}(\Gamma^{ij}J_{ij}+6\Gamma^{789})\Psi \Big]\,,\\
{}\\
\L_{2}=-\,\half\,\tr_{\!{\scriptscriptstyle
N}}\Big[(\textstyle{\frac{1}{6}})^{2}(\Phi_{5}^{\,2}
+\Phi_{6}^{\,2})+(\textstyle{\frac{1}{3}})^{2}
(\Phi_{7}^{\,2}+\Phi_{8}^{\,2}+\Phi_{9}^{\,2})\Big]\,,
\ea\label{LM5LLL}
\end{equation}
where\footnote{Ten dimensional gamma matrices are in
our convention,  $\left(\ba{ll}0&\tilde{\Gamma}^{{\scriptscriptstyle M}}\\
\Gamma^{{\scriptscriptstyle M}}&0\ea\right)$,
$\tilde{\Gamma}^{{\scriptscriptstyle
M}}=\Gamma_{{\scriptscriptstyle M}}$, $M=0,1,2,\cdots,9$.}
$i=1,2,3,4$, $a=5,6,7,8,9$, $r=7,8,9$, $\Gamma^{0}=-1$, and our
choice of the metric  for the five dimensional Minkowskian
spacetime is $\eta=\mbox{diag}(-++++)$.
Any product is to be understood as   the non-commutative star product. \\

The supersymmetry transformations are from (\ref{susytr})
\begin{equation}
\ba{l} \delta A_{\mu}=i\Psi^{\dagger}\Gamma_{\mu}\E(t)\,,~~~~~~~~~~~\delta \Phi_{a}=i\Psi^{\dagger}\Gamma_{a}\E(t)\,,\\
{}\\
\delta\Psi=\Big[\half
F_{\mu\nu}\tilde{\Gamma}^{\mu}\Gamma^{\nu}+D_{\mu}\Phi_{a}\tilde{\Gamma}^{\mu}\Gamma^{a}-i\half
[\Phi_{a},\Phi_{b}]\Gamma^{ab}-\textstyle{\frac{\mu}{12}}(\Phi_{a}\Gamma^{a}\Gamma^{789}+3\Gamma^{789}\Phi_{a}\Gamma^{a})\Big]\E(t)\\
{}\\
~~~~~~~~+\left[\ba{l}-\half\theta^{-1}_{\,ij}\Gamma^{ij}
+\textstyle{\frac{\mu}{6}}\left((\theta^{-1}_{\,1i}x^{i}+A_{1})\Gamma^{1}
+(\theta^{-1}_{\,2i}x^{i}+A_{2})\Gamma^{2}\right)(\Gamma^{789}-\Gamma^{12})\\
{}\\
+\textstyle{\frac{\mu}{6}}\left((\theta^{-1}_{\,3i}x^{i}+A_{3})\Gamma^{3}
+(\theta^{-1}_{\,4i}x^{i}+A_{4})\Gamma^{4}\right)(\Gamma^{789}-\Gamma^{34})\ea\right]\E(t)\,,
\label{LM5susytr}\ea
\end{equation}
where
\begin{equation}
\ba{cc}
\displaystyle{\E(t)=e^{\frac{\mu}{12}(-\Gamma^{12}-\Gamma^{34}+\Gamma^{789})t}\E\,,}~~~&~~~
\E={C}\E^{\ast}\,.\ea
\end{equation}
Thus the full supersymmetry remains unbroken for this
reformulation, which is no surprise as the non-commutative five dimensional
action (\ref{LM5action}) is merely  a particular manifestation of
the background independent
M-theory matrix model. \\

In the next section  by taking the commutative limit,
$\theta^{ij}\rightarrow 0$  while  keeping $\g$  fixed, we obtain
the worldvolume action for the longitudinal five  branes on the
pp-wave without the stacks of the D2 branes, as their charge densities
become
\begin{equation}
\ba{cc} (l_{p}^{6}/R^{3})[X^{i},X^{j}]=\g^{-2}{\cal
O}(\theta)\rightarrow
0\,,~~&~~(l_{p}^{6}/R^{3})\epsilon_{0ijkl}X^{i}X^{j}X^{k}X^{l}=\g^{-2}\times\mbox{const}\,.
\ea
\end{equation}
In particular we will see that the dynamical supersymmetry
reduces from sixteen to four.\footnote{Note that in field
theories, contrary to the one dimensional matrix model, the
kinematical supersymmetry is not physical at the quantum level,
since the relevant  supercharge would diverge with the space
volume factor.}\newpage

\section{Worldvolume action for the longitudinal five  branes on a pp-wave\label{WV}}
\subsection{Commutative   five dimensional   ${{\cal N}\!=\!1/2}$ worldvolume action}
Taking the commutative limit,\footnote{Due to the
anti-self-duality, there is essentially only one parameter to take
the limit.} $\theta^{ij}\rightarrow 0$, while keeping $\g$  fixed,
we first observe that the supersymmetry transformation of the
fermions (\ref{LM5susytr}) becomes singular. To remedy the
problem one should impose the following constraint on the Killing
spinor,
\begin{equation}
\Gamma^{12}\E=\Gamma^{34}\E=\Gamma^{789}\E\,,\label{EEG}
\end{equation}
which also  implies,  with   the anti-self-duality,
$\theta^{-1}_{\,ij}\Gamma^{ij}\E=0$. The constraint is in fact
equivalent to
\begin{equation}
\Omega\E=\E\,,\label{OmegaEE}
\end{equation}
where $\Omega$ is   the $4/16$ projection matrix given in
(\ref{Omega}). Hence the unbroken supersymmetry of the
longitudinal five  branes reappear precisely  as the supersymmetry
of the worldvolume theory. In the commutative limit where the
star product is replaced by the ordinary product, the action is
of the  same form as (\ref{LM5action},\,\ref{LM5LLL}), namely five dimensional
super Yang-Mills-K\"{a}hler-Chern-Simons-Myers action with four
supersymmetries.  The supersymmetry transformations reduce
to\footnote{Direct manipulation in the commutative setup indeed
shows  that the above supersymmetry transformations subject to
the constraint (\ref{constraintE}) leave the action invariant.}
\begin{equation}
\ba{l} \delta A_{\mu}=i\Psi^{\dagger}\Gamma_{\mu}\E(t)\,,~~~~~~~~~~~\delta \Phi_{a}=i\Psi^{\dagger}\Gamma_{a}\E(t)\,,\\
{}\\
\delta\Psi=\Big[\half
F_{\mu\nu}\tilde{\Gamma}^{\mu}\Gamma^{\nu}+D_{\mu}\Phi_{a}\tilde{\Gamma}^{\mu}\Gamma^{a}-i\half
[\Phi_{a},\Phi_{b}]\Gamma^{ab}
-\textstyle{\frac{\mu}{12}}(\Phi_{a}\Gamma^{a}\Gamma^{789}+3\Gamma^{789}\Phi_{a}\Gamma^{a})\Big]\E(t)\,,\ea
\end{equation}
where $a=5,6,7,8,9$ and
\begin{equation}
\ba{ccc}
\displaystyle{\E(t)=e^{-\frac{\mu}{12}\Gamma^{789}t}\E\,,}~~~&~~~
\E={C}\E^{\ast}\,,~~~&~~~\Omega\E=\E\,.\ea\label{constraintE}
\end{equation}
As the five dimensional Lorentz symmetry is explicitly broken,
the supersymmetry can be half of the ``minimal" one, or ``${\cal N}=1/2$".\\

At this point, it is interesting   to compare  with the ordinary
BFSS matrix model  or the $\mu=0$ case. In that case, the only
singular piece in the $\theta^{ij}\rightarrow 0$ limit of  the
supersymmetry transformation is
$-\half\theta^{-1}_{\,ij}\Gamma^{ij}\E$. Unlike the  $\mu\neq 0$
case,  this singularity can be removed by  the kinematical
supersymmetry transformations, as both the dynamical and
kinematical  supersymmetry transformations do not have the
explicit time dependency   when $\mu=0$. Thus in the $\mu=0$ case
the full dynamical  supersymmetry remains unbroken in the
commutative limit.
Nevertheless, both in the $\mu=0$ and $\mu\neq 0$ cases, the
commutative worldvolume actions are equipped with the same number
of supersymmetries the longitudinal five  branes preserve, i.e. 16
for $\mu=0$ and 4 for $\mu\neq 0$.\\

From (\ref{OmegaEE}) it follows $\Gamma^{56}\E=\Gamma^{789}\E$ in
addition to (\ref{EEG}). Thus, if we redefine the fermions and two
of the Higgs, using the time dependent $\mbox{SO}(2)$ rotation,
\begin{equation}
\ba{lll}
\Psi\rightarrow\displaystyle{e^{-\frac{\mu}{12}\Gamma^{56}t}\Psi\,,}~&\Phi_{5}\rightarrow
\cos(\mu t/6)\Phi_{5}-\sin(\mu
t/6)\Phi_{6}\,,~&\Phi_{6}\rightarrow \cos(\mu
t/6)\Phi_{6}+\sin(\mu t/6)\Phi_{5}\,,\ea\label{SO2}
\end{equation}
the explicit time dependency in the supersymmetry transformations
will disappear.\footnote{Note that the direction of the rotation is
opposite to (\ref{XROTATION},\,\ref{PSIROTATION}).}  In terms of
the new variables, our ${{\cal N}\!=\!1/2}$  super
Yang-Mills-K\"{a}hler-Chern-Simons-Myers   action for  the
description of the  longitudinal five  branes  on the pp-wave becomes
\begin{equation}
\ba{ll} {\cal S}=\displaystyle{\frac{1}{\g^{2}} \int\,dx^{5}}~
\L_{0}+\mu\L_{1}+\mu^{2}\L_{2}\,,~~~&~~~~\displaystyle{\g=\sqrt{R}}\,,
\ea
\end{equation}

\begin{equation}
\ba{ll}\multicolumn{2}{l}{\L_{0}=\tr_{\!{\scriptscriptstyle
N}}\Big[-\quarter F_{\mu\nu}F^{\mu\nu}-\half
D_{\mu}\Phi_{a}D^{\mu}\Phi_{a}+\quarter
[\Phi_{a},\Phi_{b}]^{2}-i\half\Psi^{\dagger}\Gamma^{\mu}D_{\mu}\Psi-\half\Psi^{\dagger}\Gamma^{a}[\Phi_{a},\Psi]\Big]\,,}\\
{}&{}\\
\L_{1}=\tr_{\!{\scriptscriptstyle
N}}\Big[&\textstyle{\frac{1}{12}}\epsilon^{\lambda\mu\nu
ij}\Tr(A_{\lambda}\partial_{\mu}A_{\nu}-\textstyle{i\frac{2}{3}}A_{\lambda}A_{\mu}A_{\nu})J_{ij}
+\textstyle{\frac{1}{6}}\epsilon^{pq}\Phi_{p}D_{0}\Phi_{q}
-i\textstyle{\frac{1}{3}}\epsilon^{rst}\Phi_{r}\Phi_{s}\Phi_{t}\\
{}&{}\\
{}&~+i\textstyle{\frac{1}{24}}\Psi^{\dagger}(\Gamma^{12}+\Gamma^{34}-\Gamma^{56}+3\Gamma^{789})\Psi \Big]\,,\\
{}&{}\\
\multicolumn{2}{l}{\L_{2}=-\,\half\,\tr_{\!{\scriptscriptstyle
N}}\Big[(\textstyle{\frac{1}{3}})^{2}(\Phi_{7}^{\,2}+\Phi_{8}^{\,2}+\Phi_{9}^{\,2})\Big]\,,}
\ea\label{WVLM5LLL}
\end{equation}
where $i=1,2,3,4$, $a=5,6,7,8,9$, $p=5,6$, $r=7,8,9$, and
$\epsilon^{01234}=\epsilon^{56}=\epsilon^{789}=1$. \\

The supersymmetry transformations are
\begin{equation}
\ba{l}\delta A_{\mu}=i\Psi^{\dagger}\Gamma_{\mu}\E\,,~~~~~~~~~~~
\delta \Phi_{a}=i\Psi^{\dagger}\Gamma_{a}\E\,,\\
{}\\
\delta\Psi=\Big[\half
F_{\mu\nu}\tilde{\Gamma}^{\mu}\Gamma^{\nu}+D_{\mu}\Phi_{a}\tilde{\Gamma}^{\mu}\Gamma^{a}-i\half
[\Phi_{a},\Phi_{b}]\Gamma^{ab}
+\textstyle{\frac{\mu}{3}}(\Phi_{p}\Gamma^{p}-\Phi_{r}\Gamma^{r})\Gamma^{789}
\Big]\E\,,\ea\label{WVsusytr}
\end{equation}
where $\E$ is  a time independent constant  spinor subject to
$\E={C}\E^{\ast}$ and
$\Omega\E=\E$.   Note that now the supersymmetry transformations do not have the explicit time dependency, which implies that the supercharges commute with the Hamiltonian.\\

For the later reference, we give the equations of motion,
\begin{equation}
\ba{l} D_{\nu}F^{\nu}{}_{0}+i[\Phi_{a},D_{0}\Phi_{a}]
+\half\{\Psi^{\dagger}{}^{\alpha},\Psi_{\alpha}\}
-\textstyle{\frac{\mu}{3}}(F_{12}+F_{34}+i[\Phi_{5},\Phi_{6}])=0\,,\\
{}\\
D_{\nu}F^{\nu}{}_{i}+i[\Phi_{a},D_{i}\Phi_{a}]
+\half\{\Psi^{\dagger}{}^{\alpha},(\Gamma_{i}\Psi)_{\alpha}\}
+\textstyle{\frac{\mu}{3}}J_{ij}F_{j0}=0\,,\\
{}\\
D_{\mu}D^{\mu}\Phi_{p}-[\Phi_{a},[\Phi_{a},\Phi_{p}]]
+\half\{\Psi^{\dagger}{}^{\alpha},(\Gamma_{p}\Psi)_{\alpha}\}+\textstyle{\frac{\mu}{3}}\epsilon_{pq}D_{0}\Phi_{q}=0\,,\\
{}\\
D_{\mu}D^{\mu}\Phi_{r}-[\Phi_{a},[\Phi_{a},\Phi_{r}]]
+\half\{\Psi^{\dagger}{}^{\alpha},(\Gamma_{r}\Psi)_{\alpha}\}
-i\mu\epsilon_{rst}\Phi_{s}\Phi_{t}-(\textstyle{\frac{\mu}{3}})^{2}\Phi_{r}=0\,,\\
{}\\
\Gamma^{\mu}D_{\mu}\Psi-i\Gamma^{a}[\Phi_{a},\Psi]
-\textstyle{\frac{\mu}{12}}(\Gamma^{12}+\Gamma^{34}-\Gamma^{56}+3\Gamma^{789})\Psi=0\,.
\ea\label{EOM}
\end{equation}
\newpage

\subsection{Supersymmetry algebra}
The Noether charge of the supersymmetry can be written in terms of
the supercharge and the supersymmetry parameter as
\begin{equation}
\displaystyle{i\!\int\!
dx^{4}\,\trN\Big(\Psi^{\dagger}\delta\Psi\Big)=iQ^{\dagger}\E=-i\E^{\dagger}Q\,.}
\label{Noether}
\end{equation}
The  supercharge is explicitly of the form, with $a=5,6,7,8,9$,
\begin{equation}
Q=\Omega\displaystyle{\int dx^{4}\,\trN\Big[-\half
F_{\mu\nu}\Gamma^{\mu}\tilde{\Gamma}^{\nu}+D_{\mu}\Phi_{a}\Gamma^{a}\tilde{\Gamma}^{\mu}+i\half
[\Phi_{a},\Phi_{b}]\Gamma^{ab}
+\textstyle{\frac{\mu}{3}}\Phi_{a}\Gamma^{a}\Gamma^{789}
\Big]\Psi}\,,
\end{equation}
and satisfy
\begin{equation}
\ba{ll}Q=C(Q^{\dagger})^{T}\,,~~~&~~~ Q=\Omega Q\,.\ea
\end{equation}

The supersymmetry algebra of the five dimensional  ${\cal N}=1/2$ worldvolume
theory is found to be, after some tedious manipulation, ({\it cf.} \cite{Lee:2002vx})
\begin{equation}
[H,Q]=0\,,
\end{equation}

\begin{equation}
\ba{ll}
{}[M_{56},Q]=i\textstyle{\frac{1}{2}}\Gamma_{56}Q\,,~~~~~&~~~~~
{}[M_{rs},Q]=i\textstyle{\frac{1}{2}}\Gamma_{rs}Q\,,\\
{}&{}\\
{}[M_{r},M_{s}]=i\epsilon_{rst}M_{t}\,,~~~~~~~&~~~~~~M_{r}=\half\epsilon_{rst}M_{st}\,,\ea
\end{equation}

\begin{equation}
{}\{Q,\,Q^{\dagger}\}=2\Omega\Big[H-\R-\textstyle{\frac{\mu}{3}}
M_{56}+\Gamma^{r}(\R_{r}+\textstyle{\frac{\mu}{3}} M_{r})
+\Gamma^{135r}\A_{r}+\Gamma^{246r}\B_{r}\Big]\Omega\,.\label{susyalge}
\end{equation}
Here $H$ is the Hamiltonian of which the bosonic part reads
\begin{equation}
H=\displaystyle{\int dx^{4}\,\trN\Big[\half F_{0i}^{2}+\quarter
F_{ij}^{2}+\half D_{0}\Phi_{a}^{2}+\half
D_{i}\Phi_{a}^{2}-\quarter[\Phi_{a},\Phi_{b}]^{2}
+i\textstyle{\frac{\mu}{3}}\epsilon^{rst}\Phi_{r}\Phi_{s}\Phi_{t}+\half(\textstyle{\frac{\mu}{3}})^{2}\Phi_{r}^{2}\Big]\,,}
\end{equation}
$M_{56}$, $M_{rs}$ are $\so(2)$, $\so(3)$ $R$-symmetry generators,
\begin{equation}
\ba{l} M_{56}=\displaystyle{\int
dx^{4}\,}\trN\Big[\epsilon^{pq}D_{0}\Phi_{p}\Phi_{q}-\textstyle{\frac{\mu}{6}}(\Phi_{5}^{2}+
\Phi_{6}^{2})-i\textstyle{\frac{1}{4}}\Psi^{\dagger}\Gamma_{56}\Psi\Big]\,,\\
{}\\
M_{rs}=\displaystyle{\int
dx^{4}\,}\trN\Big[D_{0}\Phi_{r}\Phi_{s}-D_{0}\Phi_{s}\Phi_{r}-i\textstyle{\frac{1}{4}}\Psi^{\dagger}\Gamma_{rs}\Psi\Big]\,,\ea
\end{equation}
and $\R$, $\R_{r}$, $\A_{r}$, $\B_{r}$  are  real central charges
given by the boundary terms,
\begin{equation}
\ba{l} \R=\half\displaystyle{\int
dx^{4}\,\partial_{i}\,\trN\Big[J^{ij}\epsilon^{pq}\Phi_{p}D_{j}\Phi_{q}
-\epsilon^{0ijkl}(A_{j}\partial_{k}A_{l}-i\textstyle{\frac{2}{3}}A_{j}A_{k}A_{l})\Big]}\,,\\
{}\\
\R_{r}=\half\epsilon_{rst}\displaystyle{\int
dx^{4}\,\partial_{i}\,\trN\left[J^{ij}D_{j}\Phi_{s}\Phi_{t}\right]\,,}\\
{}\\
\A_{r}=\displaystyle{\int
dx^{4}\,\partial_{i}\,\trN\Big[h^{ij}\Phi_{r}(D_{j}\Phi_{5}-J_{jk}D_{k}\Phi_{6})\Big]\,,}\\
{}\\
\B_{r}=-\displaystyle{\int
dx^{4}\,\partial_{i}\,\trN\Big[h^{ij}\Phi_{r}(D_{j}\Phi_{6}+J_{jk}D_{k}\Phi_{5})\Big]\,,}\ea
\end{equation}
where we set $h^{31}=h^{24}=-h^{13}=-h^{42}=1$ and others zero.
Note that $\R$ contains the Chern-Pontryagin density, $F\wedge
F$, which counts the number of D0 branes dissolved in
the longitudinal five  branes. For other central charges, we do not have
clear interpretations yet in terms of the extended objects in the string
theory. \\

The numbers of degrees in the left and right hand sides of
(\ref{susyalge}) match as
\begin{equation}
10=1+3+3+3\,,
\end{equation}
as $\Omega$, $\Omega\Gamma^{r}\Omega$,
$\Omega\Gamma^{135r}\Omega$, $\Omega\Gamma^{246r}\Omega$ are the
only allowed independent gamma matrix products to appear on
the right. \\

It is interesting to note that  the  spatial translation  and the
isometry of the K\"{a}hler form, $\mbox{SU}(2)$, are not part of
the ${\cal N}=1/2$ supersymmetry algebra, though they are not
broken. After all, ${\cal N}=1/2$ supersymmetry is too small to
capture all the symmetries in the model.  Compared to the
supersymmetry algebra of the BMN matrix model
\cite{Berenstein:2002jq,Hyun:2002cm,JhpMsusy,Sugiyama:2002rs},
the coefficient of $M_{56}$ appearing in the anti-commutator of
the supercharges is doubled from ${\mu/6}$ to
${\mu/3}$. This reflects  our  redefinition of $\Phi_{5}$, $\Phi_{6}$  by the rotating ones (\ref{SO2}).\\

From the positive definity, we have the following energy bound,
\begin{equation}
H\geq
\R+\textstyle{\frac{\mu}{3}}M_{56}+\left|(\hat{e}_{1})_{r}(\R_{r}+\textstyle{\frac{\mu}{3}}
M_{r})\right|+\left|(\hat{e}_{2})_{r}\A_{r}\right|+\left|(\hat{e}_{3})_{r}\B_{r}\right|\,,\label{Ebound}
\end{equation}
where $\hat{e}_{1}$, $\hat{e}_{2}$, $\hat{e}_{3}$ form an
arbitrary orthonormal real basis for the  ``$7,8,9$'' space so
that $(\hat{e}_{1})_{r}\Gamma^{r}$,
$(\hat{e}_{2})_{r}\Gamma^{135r}$,
$(\hat{e}_{3})_{r}\Gamma^{246r}$ can be simultaneously
diagonalized with the eigenvalues, $\pm 1$.

\subsection{BPS equations for the fully supersymmetric  configurations}
In this subsection we consider the BPS configurations which
preserve all the four supersymmetries. In the conventional
supersymmetric models, such fully supersymmetric configurations
would be vacua, but in the present case, the novel structure of
the supersymmetry algebra allows nontrivial fully supersymmetric
BPS configurations. They have the energy saturation,
\begin{equation}
H=\R+\textstyle{\frac{\mu}{3}}M_{56}\,,
\end{equation}
while other central and $R$-symmetry charges  vanish, $\R_{r}=\A_{r}=\B_{r}=M_{r}=0$.\\

The corresponding BPS equations can be obtained either by writing
$H-\R-\textstyle{\frac{\mu}{3}}M_{56}$ as a sum of squares or
from the supersymmetry transformation of the fermions
(\ref{deltaPsi}),
\begin{equation}
\ba{ll}
F_{0\mu}=D_{0}\Phi_{r}=0\,,~~~&~~~D_{0}\Phi_{p}-\textstyle{\frac{\mu}{3}}\epsilon_{pq}\Phi_{q}=0\,,\\
{}&{}\\
{}[\Phi_{r},\Phi_{s}]-i\textstyle{\frac{\mu}{3}}\epsilon_{rst}\Phi_{t}=0\,,~~~&~~~
D_{i}\Phi_{r}=[\Phi_{5},\Phi_{r}]=[\Phi_{6},\Phi_{r}]=0\,,\\
{}&{}\\
F_{13}+F_{42}=0\,,~~~&~~~F_{14}+F_{23}=0\,,\\
{}&{}\\
F_{12}+F_{34}-i[\Phi_{5},\Phi_{6}]=0\,,~~~&~~~D_{i}\Phi_{5}-J_{ij}D_{j}\Phi_{6}=0\,,
\ea
\end{equation}
where $i=1,2,3,4$, $p=5,6$,  $r=7,8,9$, and the BPS equations
themselves satisfy the Gauss constraint.  In particular, the
$\so(2)$ $R$-symmetry charge becomes
\begin{equation}
M_{56}=\textstyle{\frac{\mu}{6}}\displaystyle{\int
dx^{4}\,}\trN\left(\Phi_{5}^{2}+\Phi_{6}^{2}\right)\,.
\end{equation}
These BPS equations are the same as  the BPS equations, (\ref{MBPS}),
 in the original M-theory matrix
model up to the field redefinition (\ref{SO2}). After all,
the flat longitudinal five  brane is just a
particular solution of the latter and the BPS equations above in
the worldvolume theory can be interpreted  as the constraint for
the D0 branes  dissolved in the five  branes which still  preserve
the four supersymmetries.\\

The last four BPS equations are essentially identical to  the
BPS equations in Euclidean six dimensional super Yang-Mills theory
\cite{jhpBPS}. When
all the Higgs are turned off, the BPS equations reduce to the well
known anti-self-dual equations for the field strength,
$F_{ij}+\half\epsilon_{ijkl}
F_{kl}=0$, for which the ADHM construction provides the general
solutions.
On the other hand, just like in the BMN matrix model,
the classical supersymmetric vacua are given by the
constant fuzzy spheres,
\begin{equation}
\ba{cc}
[\Phi_{r},\Phi_{s}]=i\textstyle{\frac{\mu}{3}}\epsilon_{rst}\Phi_{t}\,,~~~~&~~~~
\Phi_{5}=\Phi_{6}=F_{\mu\nu}=0\,.
\ea
\end{equation}

\subsection{Energy spectra and supermultiplets}
In order to clarify the supermultiplet contents, we investigate the energy
spectra of the bosons and fermions. This can be done by
considering the
equations of motion (\ref{EOM}) for the free
or $\mbox{U}(1)$ case.\\

For the fermions, if we consider the plane wave solution, $\Psi(x)=\psi_{k}e^{ik{\cdot x}}$,  the equation of motion  becomes
\begin{equation}
\left[i\Gamma^{\mu}k_{\mu}-\textstyle{\frac{\mu}{12}}(\Gamma^{12}+\Gamma^{34}-\Gamma^{56}+3\Gamma^{789})\right]\psi_{k}=0\,.
\end{equation}
To admit a nontrivial solution it is necessary  to
impose\footnote{In the manipulation of the determinant we used the
explicit representation of the gamma matrices given in the appendix.}
\begin{equation}
\det\!\left[i\Gamma^{\mu}k_{\mu}-\textstyle{\frac{\mu}{12}}(\Gamma^{12}+\Gamma^{34}-\Gamma^{56}+3\Gamma^{789})\right]=
\left(k^{2}+(\textstyle{\frac{\mu}{3}})^{2}\right)^{2}\left(k^{2}+\textstyle{\frac{\mu}{3}}k_{0}\right)\left(k^{2}
-\textstyle{\frac{\mu}{3}}k_{0}\right)=0\,.\label{det}
\end{equation}
Thus, the fermions have the following three energy spectra,
\begin{equation}
\ba{ccc} E_{{\bf k}}=\sqrt{(\textstyle{\frac{\mu}{3}})^{2}+{{\bf
k}}^{2}\,}\,,~~~&~~~ E^{+}_{{\bf
k}}=\sqrt{(\textstyle{\frac{\mu}{6}})^{2}+{{\bf
k}}^{2}\,}+\textstyle{\frac{|\mu|}{6}}\,,~~~&~~~ E^{-}_{{\bf
k}}=\sqrt{(\textstyle{\frac{\mu}{6}})^{2}+{{\bf
k}}^{2}\,}-\textstyle{\frac{|\mu|}{6}}\,.\ea
\end{equation}
For each spectrum there are  four, two and  two fermionic modes,
respectively.\\

 Similarly one can obtain the energy spectra for the
bosons. The gauge fields consist of three independent modes having
the above three energy spectra, $E_{{\bf k}}, E^{+}_{{\bf
k}},E^{-}_{{\bf k}}$, respectively. The Higgs fields,
$\Phi_{5}, \Phi_{6}$, decompose into
two modes which have the energy spectra, $E^{+}_{{\bf
k}},E^{-}_{{\bf k}}$,
while the other three Higgs fields, $\Phi_{r}$, have only one
spectrum, $E_{{\bf k}}$.\\

In fact, the $\pm \frac{|\mu|}{6}$ factors in $E^{\pm}_{{\bf k}}$
are the reminiscent of  the coordinate transformations  using the
$\mbox{SO}(2)$ rotations (\ref{XROTATION},\ref{SO2}). They
coincide with the frequencies of the rotations.\\

{The energy spectra of the bosons and fermions are summarized  in Table \ref{T1}.
\begin{table}[htb]
\begin{center}
\begin{tabular}{l|cccccc}
~~~energy spectra~~~~~~~~ & $~~~~\Psi~~~~$ & $~~~A_{\mu}~~~$ &
$~~~\Phi_{5},\Phi_{6}~~~$ & $~~~\Phi_{7},\Phi_{8},\Phi_{9}~~~$\\
 \hline
{}&{}&{}&{}\\
$E_{{\bf k}}=\sqrt{(\textstyle{\frac{\mu}{3}})^{2}+{{\bf
k}}^{2}\,}$ & 4 & 1 & 0 & 3\\
{}&{}&{}&{}\\
$E^{+}_{{\bf k}}=\sqrt{(\textstyle{\frac{\mu}{6}})^{2}+{{\bf
k}}^{2}\,}+\textstyle{\frac{|\mu|}{6}}$ & 2 & 1 & 1 & 0\\
{}&{}&{}&{}\\
$E^{-}_{{\bf k}}=\sqrt{(\textstyle{\frac{\mu}{6}})^{2}+{{\bf
k}}^{2}\,}-\textstyle{\frac{|\mu|}{6}}~~~$ & 2 & 1 & 1 & 0\\
\end{tabular}
\end{center}
\caption{Energy spectra and the numbers of bosons and fermions.}
\label{T1}
\end{table}}

Clearly each line forms a separate supermultiplet. Note that
  it is the coefficient of the Fourier mode, the
creation or annihilation operator, not the $c$-number part,
$e^{ik{\cdot x}}$, that transforms under  the adjoint action of
the  supercharges  on the canonically quantized  fields.

\section{Conclusion\label{conclusion}}
We have obtained a five dimensional $\mbox{U}(N)$  ${\cal N}=1/2$
super Yang-Mills action with  the K\"{a}hler-Chern-Simons
term and   the Myers term  as a low energy worldvolume description of the
longitudinal five  branes  in a maximally  supersymmetric pp-wave background.\\

We derived the action utilizing  the known rotating longitudinal
five  brane solution preserving four supersymmetries in the BMN
matrix model.  Adopting the inertial or comoving frame, we
reformulate  the  matrix model in a new coordinate system which
involves the replacement  of  some bosonic mass terms by the
``Chern-Simons'' term and the modification of the fermion's mass
term. In this setup the `flat' and `static' longitudinal five  brane
solution was identified and shown to preserve four dynamical
supersymmetries. Expanding the modified matrix model around the
solution, we first obtained a non-commutative field theory
naturally equipped with the full sixteen dynamical
supersymmetries. In the commutative limit, we showed only four
supersymmetries survive, resulting in the ${\cal
N}=1/2$ model. \\

In the original BMN matrix model which is written in the maximally
symmetric coordinate system, the longitudinal five branes should  rotate in order to
preserve the supersymmetries.  The K\"{a}hler structure in the worldvolume action
is inherited from the rotating directions of the longitudinal five branes.
In this sense, due to the presence of the K\"{a}hler form, the five dimensional
Lorentz symmetry is spontaneously broken. This accounts the emergence of the half of
the ``minimal'' supersymmetry in five dimensions. \\

We wrote the supersymmetry algebra explicitly,  identifying all
the possible  central charges. Thanks to the novel structure of
the algebra, the ${\cal N}=1/2$ model admits the BPS
configurations which preserve all the four supersymmetries.  In
particular, when all the Higgs fields  are turned off, they
reduce to the ordinary anti-self-dual equations for the field
strength,
while the classical supersymmetric vacua are given by the constant fuzzy spheres.\\

We obtained the novel energy spectra of the bosons and fermions in the
worldvolume action. The results show that the model contains
three different supermultiplets embedded in a nontrivial way.\\

The resulting worldvolume action possesses four supersymmetries,
which is natural as we  started with the   five  brane
configuration  preserving four dynamical supersymmetries in the
matrix model. According to  the classification of the BPS
equations in the BMN matrix model \cite{Park:2002cb}, it appears
that there is  no flat longitudinal five  brane configuration
which preserves other than four dynamical supersymmetries. This
is certainly  true within the matrix formulation setup of the
M-theory. However, recently it was shown that the mass
deformation of the  DLCQ matrix model for the longitudinal five 
branes \cite{Aharony} is possible, while keeping eight dynamical
supersymmetries \cite{Kim:2002cr}. This might suggest that, just
like the transverse five  branes, more supersymmetric longitudinal
five branes may exist in the M-theory on the pp-wave, which the
matrix model can not capture.\\

Contrary to  the D-brane worldvolume actions  in the  flat
background, the conventional dimensional reduction of the present
five dimensional ${\cal N}=1/2$ action would not correspond to
the T-duality of string theory due to the nontrivial pp-wave geometry.
The worldvolume actions for other branes on the pp-wave should be
obtained case by case.

\acknowledgments{The authors wish to thank  Kimyeong Lee, Jae-Suk
Park,  Soo-Jong Rey and Sang-Heon Yi  for the enlightening
discussions. The work of SH was supported in part by grant No. R01-2000-00021 from the Basic Research Program of the Korea Science and Engineering Foundation.}
\newpage
\appendix
\section{Appendix}
A useful identity to derive the BPS equations preserving all the
four supersymmetries in the worldvolume action is
\begin{equation}
\ba{l}\Big[\half
F_{\mu\nu}\tilde{\Gamma}^{\mu}\Gamma^{\nu}+D_{\mu}\Phi_{a}\tilde{\Gamma}^{\mu}\Gamma^{a}-i\half
[\Phi_{a},\Phi_{b}]\Gamma^{ab}
+\textstyle{\frac{\mu}{3}}(\Phi_{p}\Gamma^{p}-\Phi_{r}\Gamma^{r})\Gamma^{789}
\Big]\Omega\\
{}\\
=F_{0i}\Gamma^{i}+D_{0}\Phi_{r}\Gamma^{r}+D_{i}\Phi_{r}\Gamma^{ir}+
(D_{0}\Phi_{p}-\textstyle{\frac{\mu}{3}}\epsilon_{pq}\Phi_{q})\Gamma^{p}
-i\half([\Phi_{r},\Phi_{s}]-i\textstyle{\frac{\mu}{3}}\epsilon_{rst}\Phi_{t})\Gamma^{rs}\\
{}\\
~~~~~-i[\Phi_{p},\Phi_{r}]\Gamma^{pr}
+(D_{i}\Phi_{5}-J_{ij}D_{j}\Phi_{6})\Gamma^{i5}+(F_{13}+F_{42})\Gamma^{13}+(F_{14}+F_{23})\Gamma^{14}\\
{}\\
~~~~~+(F_{12}+F_{34}-i[\Phi_{5},\Phi_{6}])\Gamma^{12}\,,\ea\label{deltaPsi}
\end{equation}
where $i=1,2,3,4$, $p=5,6$, $r=7,8,9$.\\

Evaluating the anti-commutator of the supercharges to derive the
supersymmetry algebra (\ref{susyalge}), one needs the following
Fierz identities for the nine dimensional gamma matrices,
$(\Gamma^{A})_{{\alpha}}{}^{{\beta}}$,
${\alpha},{\beta}=1,2,\cdots,16$,
\begin{equation}
\ba{l}
\delta^{{\alpha}}{}_{{\gamma}}\delta^{{\beta}}{}_{{\delta}}-\delta^{{\alpha}}{}_{{\delta}}\delta^{{\beta}}{}_{{\gamma}}
=\textstyle{\frac{1}{16}}(C^{-1}\Gamma^{AB})^{{\alpha}{\beta}}(\Gamma_{AB}C)_{{\gamma}{\delta}}+
\textstyle{\frac{1}{48}}(C^{-1}\Gamma^{ABC})^{{\alpha}{\beta}}(\Gamma_{ABC}C)_{{\gamma}{\delta}}\,,\\
{}\\
(\Gamma^{AB})_{{\alpha}}{}^{{\gamma}}(C^{-1}\Gamma_{B})^{{\beta}{\delta}}
+(C^{-1}\Gamma^{AB})^{{\beta}{\delta}}(\Gamma_{B})_{{\alpha}}{}^{{\gamma}}\,+\,({\gamma}\leftrightarrow{\delta})=
2(\Gamma^{A})_{{\alpha}}{}^{{\beta}}C^{-1}{}^{{\gamma}{\delta}}
-2\delta_{{\alpha}}{}^{{\beta}}(C^{-1}\Gamma^{A})^{{\gamma}{\delta}}\,.
\ea \label{APP1}
\end{equation}

In the manipulation of the determinant (\ref{det}),   we used the
following  representation of  the ``ten'' dimensional gamma
matrices,
\begin{equation}
\begin{array}{lll}
\Gamma^{0}=-1\otimes
1\,,~~&~~\Gamma^{m}=1\otimes\gamma^{m}\,,~~&~~\Gamma^{r}=\sigma^{r-6}\otimes\gamma^{(7)}\,,
\end{array}
\end{equation}
where $m=1,2,\cdots,6$, $~r=7,8,9$,
$~\sigma^{1},\sigma^{2},\sigma^{3}$ are the usual Pauli matrices,
and $\gamma^{m}$'s are the six dimensional gamma matrices,
\begin{equation}
\ba{ll}
\gamma^{m}=\left(\begin{array}{cc}0&\rho^{m}\\
({\rho}^{m})^{\dagger}&0\end{array}\right)\,, ~~~&~~~
\gamma^{(7)}=i\gamma^{4}\gamma^{5}\cdots\gamma^{9}=\left(\begin{array}{rr}1&0\\0&-1\end{array}\right)\,,
\ea
\end{equation}
with the  anti-symmetric $4\times 4$ matrices \cite{Park:1998nr},
\begin{equation}
\begin{array}{lll}
\rho^{1}=\left(\ba{cc}i\epsilon&0\\0&-i\epsilon^{-1}\ea\right)\,,~&~\rho^{2}=\left(\ba{cc}\epsilon&0\\0&\epsilon^{-1}\ea\right)\,,~&~
\rho^{3}=\left(\ba{cc}0&i\sigma^{3}\\-i(\sigma^{3})^{T}&0\ea\right)\,,\\
{}&{}&{}\\
\rho^{4}=\left(\ba{rr}0&1\\-1&0\ea\right)\,,~&~\rho^{5}=\left(\ba{cc}0&i\sigma^{1}\\-i(\sigma^{1})^{T}&0\ea\right)\,,~&~
\rho^{6}=\left(\ba{cc}0&i\sigma^{2}\\-i(\sigma^{2})^{T}&0\ea\right)\,.
\end{array}
\label{rhochoice}
\end{equation}

\end{document}